\documentclass[aps,prl,reprint,superscriptaddress]{revtex4-2}
\usepackage{amsmath}
\usepackage{amssymb}
\usepackage{amsfonts}
\usepackage{bbm}
\usepackage{bm}
\usepackage{graphicx}
\usepackage{color}
\usepackage{dcolumn}
\usepackage{MnSymbol}
\usepackage{times}
\usepackage{physics}
\usepackage[colorlinks=true,linkcolor=blue,citecolor=blue,urlcolor=blue]{hyperref}

\def\map{\mathcal}

\usepackage{color}
\definecolor{dred}{rgb}{.8,0.2,.2}
\definecolor{ddred}{rgb}{.8,0.5,.5}
\definecolor{dblue}{rgb}{.2,0.2,.8}
\definecolor{dgreen}{rgb}{.2,0.5,.2}

\usepackage{amsthm,amsmath,amssymb}
\usepackage{mathrsfs}

\definecolor{dred}{rgb}{.8,0.2,.2}
\definecolor{ddred}{rgb}{.8,0.5,.5}
\definecolor{dblue}{rgb}{.2,0.2,.8}
\definecolor{dgreen}{rgb}{.2,0.5,.2}

\newcommand*{\qcircuit}{$\begin{matrix}\Qcircuit @C=2em @R=2em}
\newcommand*{\eqcircuit}{\end{matrix}$}

\graphicspath{{figures/}}

\newcommand*{\physus}{Department of Physics, State Key Laboratory of Quantum Functional Materials, and Guangdong Basic Research Center of Excellence for Quantum Science, Southern University of Science and Technology, Shenzhen 518055, China}

\begin{document}

\preprint{APS/123-QED}

\title{Experimental Virtual Quantum Broadcasting}

\author{Yuxuan Zheng}
\thanks{These authors contributed equally to this work.}
\affiliation{\physus}

\author{Xinfang Nie}
\thanks{These authors contributed equally to this work.}
\email{niexf@sustech.edu.cn}

\affiliation{Quantum Science Center of Guangdong-Hong Kong-Macao Greater Bay Area, Shenzhen 518045, China}
\affiliation{\physus}

\author{Hongfeng Liu}
\thanks{These authors contributed equally to this work.}
\affiliation{\physus}

\author{Yutong Luo}
\affiliation{School of Physics, Trinity College Dublin, College Green, Dublin 2, D02 K8N4, Ireland}
\affiliation{Trinity Quantum Alliance, Unit 16, Trinity Technology and Enterprise Centre, Pearse Street, Dublin 2, D02 YN67, Ireland}

\author{Dawei Lu}
\email{ludw@sustech.edu.cn}
\affiliation{\physus}
\affiliation{Quantum Science Center of Guangdong-Hong Kong-Macao Greater Bay Area, Shenzhen 518045, China}

\author{Xiangjing Liu}
\email{xiangjing.liu@cnrsatcreate.sg}
\affiliation{CNRS@CREATE, 1 Create Way, 08-01 Create Tower, Singapore 138602, Singapore }
\affiliation{MajuLab, CNRS-UCA-SU-NUS-NTU International Joint Research Unit, Singapore}
\affiliation{Centre for Quantum Technologies, National University of Singapore, Singapore 117543, Singapore}

\date{\today}

\begin{abstract}

The quantum no-broadcasting theorem states that it is fundamentally impossible to perfectly replicate an arbitrary quantum state, even if correlations between the copies are allowed. While quantum broadcasting cannot occur through any physical process, it can be achieved via postprocessing of experimental data using a process called virtual quantum broadcasting (VQB). In this work, we report the
first experimental implementation of a quantum circuit based on the linear combination of unitaries, integrated with a post-processing protocol, to realize VQB in a nuclear magnetic resonance system. VQB can be expressed as a linear combination of two channels: the universal cloner, which broadcasts the target quantum state, and the universal antisymmetrizer, which reduces broadcasting error. We implement both channels within the same circuit and demonstrate that the universal cloner is the closest physical map to VQB. In addition, we show how the universal antisymmetrizer can be utilized to mitigate imperfections in the cloner, enabling near-ideal fidelity.  Our method is applicable to broadcasting quantum systems of any dimension.

\end{abstract}

\maketitle

\emph{\bfseries Introduction.---} Distributing quantum information to multiple recipients is of foundational importance to quantum technologies such as quantum network~\cite{wehner2018quantum}, distributed quantum computing~\cite{caleffi2024distributed,cacciapuoti2019quantum} and sensing~\cite{zhang2021distributed,Degen2017quantum}. However, quantum mechanics limits the ability to distribute quantum information, as unknown pure quantum states cannot be perfectly copied~\cite{wootters1982single}. Extending this principle to mixed states, the quantum no-broadcasting theorem states that no physical process, i.e., a completely positive trace-preserving (CPTP) map, can broadcast an unknown quantum mixed state to two separate parties while preserving its reduced density matrices~\cite{barnum1996noncommuting}.
 Nevertheless, optimal physical protocols exist that can approximately broadcast quantum states~\cite{bruss1198optimal,werner1998optimal,harrow2013church,chiribella2010quantum}.

Interestingly, quantum broadcasting can be achieved by relaxing the positivity requirement, a process referred to as virtual quantum broadcasting (VQB)~\cite{Parzygnat2024virtual}.
VQB belongs to a broader class of quantum operations known as Hermitian-preserving and trace-preserving (HPTP) maps, and is inspired by recent advancements in quantum spatiotemporal correlations~\cite{fitzsimons2015quantum,Parzygnat2023from,liu2024inferring,liu2023quantum,liu2024unification,song2024causal,fullwood2022on,jia2023quantum,fullwood2023quantum,bressanini2024quantum,Lie2024quantum,liu2024experimental,liu2024quantum,fullwood2024operator,jia2024spatiotemporal,fullwood2024general}. While infinitely many such HPTP maps satisfy the broadcasting condition, three physical assumptions uniquely single out one~\cite{Parzygnat2024virtual}. HPTP maps are linear maps that transform Hermitian operators into Hermitian operators but do not necessarily preserve positivity~\cite{PhysRevA.66.052315,PhysRevA.78.062105}. For instance, they can map a density matrix to a non-positive operator. 

HPTP maps can, however, be simulated by sampling quantum operations combined with post-processing measurement statistics from output states, aligning with a paradigm referred to as virtual operations.
Virtual operations reproduce the measurement effects necessary for quantum information processing tasks without requiring physical implementation, operating at the level of expectation values rather than directly on quantum states or their dynamics, and relying on classical post-processing of outputs~\cite{PhysRevLett.107.160401,liu2024virtual,wei2024simulating}. It has found applications in quantum error mitigation~\cite{kandala2019error,cai2023quantum,yamamoto2022error}, virtual quantum resource distillation~\cite{yuan2024virtual,takagi2024virtual,zhang2024experimental,yamamoto2024virtual} and reversing unknown quantum operations~\cite{zhu2024reserving}.
Given the fundamental significance of VQB, its explicit simulation and practical feasibility within existing quantum technologies remain open questions.

\begin{figure}[t!]
\centering
\includegraphics[width=1\linewidth]{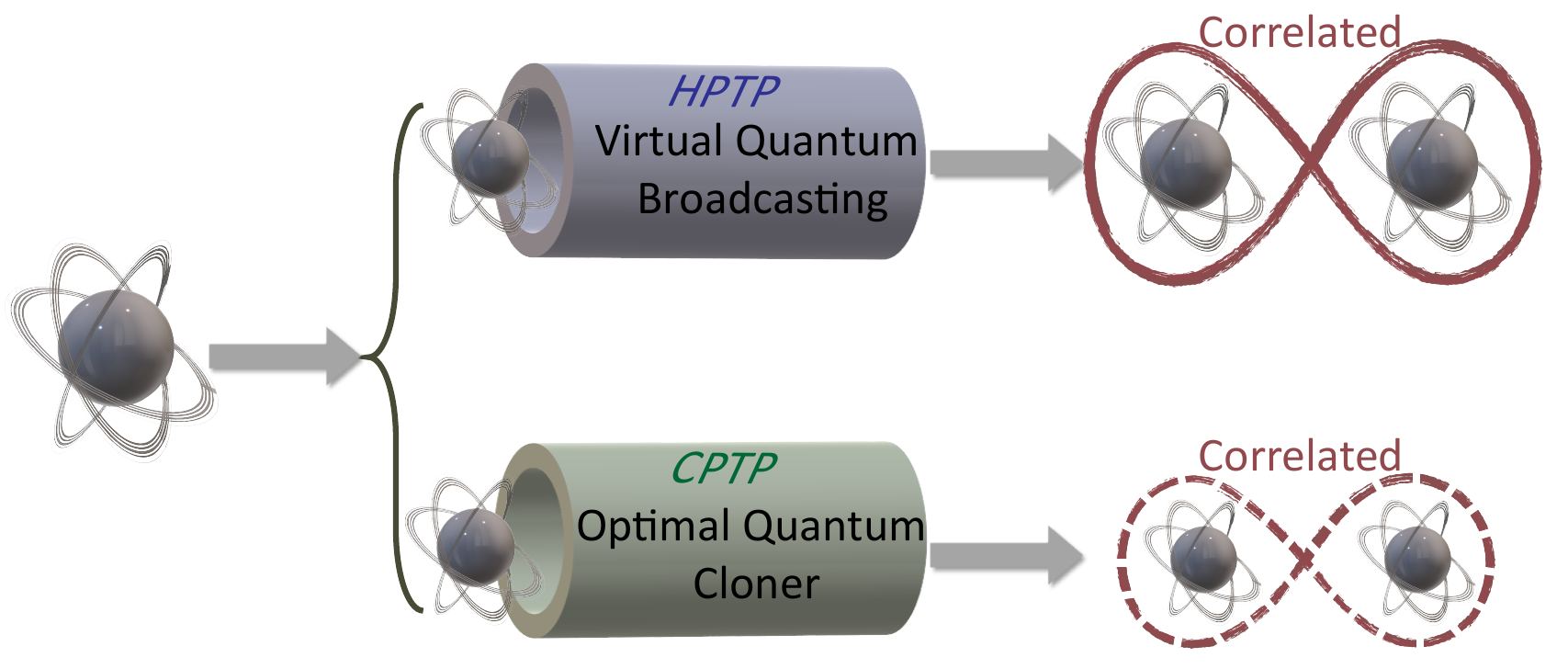}
\caption{{\bf{Illustration of the VQB and the universal cloner}}. VQB is an HPTP map that produces perfect local copies of an arbitrary quantum state while maintaining correlations between them. The universal cloner is a CPTP map that can only generate imperfect local copies of an arbitrary quantum state. The size of the qubit system represents the cloning fidelity.}
\label{Fig1}
\end{figure}

In this Letter, we employ concepts from the linear combination of unitaries~\cite{gui2006general,childs2012hamiltonian,liu2024certifying} to design an explicit protocol for implementing VQB, which is subsequently realized on a nuclear magnetic resonance (NMR) quantum processor~\cite{RevModPhys.76.1037,cory1997ensemble}. Firstly, inspired by the fact that VQB can be expressed as a linear combination of two channels—the so-called universal cloner and antisymmetrizer~\cite{Parzygnat2024virtual}—we design and implement a single quantum circuit to probabilistically realize both channels, along with a post-processing procedure to obtain VQB. Secondly, we experimentally verify that the universal cloner is the closest CPTP map to VQB by comparing the trace distances of the Choi states of the VQB map with those of a suitably chosen set of CPTP maps. Finally, we treat the imperfection of the universal cloner, i.e., the difference between the statistical outcomes of the input to the universal cloner and its local output, as the `fundamental error' imposed by quantum mechanics~\cite{wootters1982single}. We demonstrate that this error can be eliminated through classical post-processing, thereby achieving quantum broadcasting.

\emph{\bfseries Virtual quantum broadcasting, universal cloner, and its circuit.---}We briefly introduce the concept of VQB and demonstrate how to simulate it. 
A broadcasting map $\mathcal{M}$ is a linear map from a quantum system to the compound system $AB$, such that the following broadcasting condition is satisfied~\cite{barnum1996noncommuting},
\begin{align}
\Tr_{A}\map M (\rho) = \rho=\Tr_B \map M (\rho), \quad \forall\, \rho.
\label{eq:broadcasting_condition}
\end{align}
Broadcasting is not possible for CPTP maps; however, it is possible for HPTP maps. HPTP maps that satisfy Eq.~(\ref{eq:broadcasting_condition}) are termed virtual broadcasting maps and are not unique. Three requirements—1) covariance under unitary evolution, 2) invariance under the permutation of the copies, and 3) consistency with classical broadcasting—uniquely determine the so-called canonical broadcasting map~\cite{Parzygnat2024virtual},   $$\mathcal{B}(\rho) = \frac{1}{2} \{\rho_A \otimes \mathbb{I}_B, S_{AB} \}, $$
where $\{\cdot,\cdot\}$ denotes the anti-commutator, $\mathbb{I}$ is the identity matrix and $S$ is the swap operator. Interestingly, the output $\map{B}(\rho)$  is associated with a type of pseudo-density matrix~\cite{fitzsimons2015quantum,liu2023quantum}, sometimes referred to as a quantum state over time~\cite{fullwood2022on,fullwood2023quantum}, representing a quantum state undergoing trivial dynamics (see Supplemental Material (SM)~\cite{supp}). The map $\mathcal{B}$ can be decomposed into a linear combination of two CPTP maps,
\begin{align} 
\mathcal{B}= \frac{d+1}{2}\mathcal{B}^+ - \frac{d-1}{2}\mathcal{B}^-,
\end{align}
where $d$ is the dimension of systems $A$ and $B$ and $\mathcal{B}^\pm (\rho) = \frac{2}{d \pm 1} \Pi^{\pm} (\rho \otimes \mathbb{I} ) \Pi^{\pm}$ 
with $\Pi^{\pm} = \frac{1}{2}(\mathbb{I} \otimes \mathbb{I} \pm S) $. The map $\map{B}^+$ is termed the universal cloner, while $\map{B}^-$ is termed the universal anti-symmetrizer. Importantly, the map $\map{B}^+$ is shown to be the closest CPTP map to $\map{B}$ in terms of the diamond distance~\cite{Parzygnat2024virtual}. The universal cloner is constrained by the laws of quantum mechanics and cannot achieve perfect cloning fidelity ~\cite{bruss1198optimal,bruss1998optimalprl}.  A pictorial illustration of the VQB and $\map{B}^+$  is shown in Fig.~\ref{Fig1}.

\begin{figure}[tbh!]
\centering
\includegraphics[width=1\linewidth]{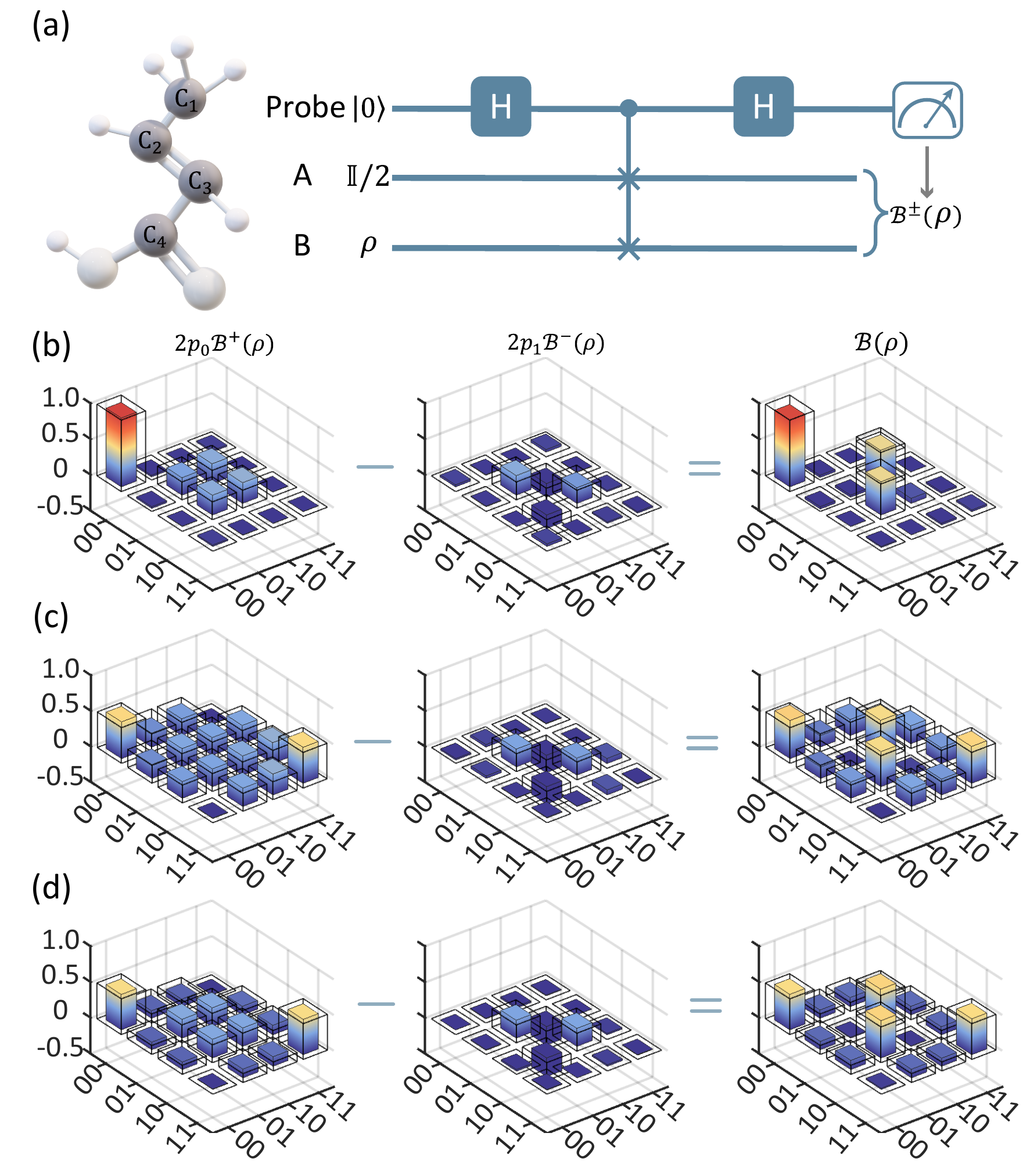}
\caption{ {\bf{Experimental results of VQB}}. (a) Molecular structure of the four-qubit quantum simulator and quantum circuit for implementing the universal cloner $\map{B}^+$ and the universal antisymmetrizer $\map{B}^-$ by projecting onto the probe qubit, with $p_0$ and  $p_1$ being the corresponding probabilities. The first three qubits are used in this quantum circuit.  $\mathcal{B}(\rho)$ is obtained by the post-processing $\map{B}^\pm(\rho)$ according to Eq.~\eqref{Eq3}. 
Results for $2p_0\map{B}^+, 2p_1\map{B}^-$ and $\map{B}(\rho)$ with three different input states:  $\ket{0}$, $\ket{+}$ and $\mathbb{I}/{4}+\ket{+}\bra{+}/2$, are shown in (b)-(d). Experimental results are represented by colored bars, while theoretical results are shown as colorless, transparent bars. 
}
\label{Fig2}
\end{figure}

The decomposition of $\map{B}$ suggests a method for its simulation. First, the two maps $\map{B}^\pm$ can be simulated using circuits designed for the linear combination of unitaries~\cite{childs2012hamiltonian,de2002experimental}, as shown in Fig.~\ref{Fig2}(a). The circuit applies for the input state $\rho$ with an any finite dimension $d$, we take $d=2$ in our experiment. Specifically, up to a normalization constant, when the control qubit is projected onto $P_0=\ket{0}\bra{0}$, the map $\map{B}^+$ is implemented; similarly, when the control qubit is projected onto $P_1=\ket{1}\bra{1}$, the map $\map{B}^-$ is implemented. Next, the linear combination of $\map{B}^+(\rho)$ and $\map{B}^-(\rho)$ can be realized through classical post-processing; for details see SM~\cite{supp}. Our approach is then implemented in a NMR quantum processor.

{\emph{\bfseries Experimental setup.---}}Our experiments are designed and conducted in an NMR setup to realize VQB and verify the optimality of $\map{B}^+$.
All the experiments are conducted on a Bruker 300 MHz spectrometer at room temperature using a four-qubit nuclear spin system composed of $^{13}$C-labeled trans-crotonic acid dissolved in $d_6$-acetone. 
The four carbon nuclear spins, labeled as C$_{1-4}$, form a four-qubit quantum processor~\cite{xi2024experimental,PhysRevLett.132.210403,PhysRevA.72.062317,PhysRevA.71.032344,PhysRevLett.86.5811}, with the molecular structure illustrated in Fig.~\ref{Fig2}(a). 
More details for experimental setup can be found in SM~\cite{supp}. 
In the NMR system, single-qubit rotations are implemented using transverse radio-frequency pulses, while two-qubit interactions are achieved through free evolution. The accuracy of experimental control can be further enhanced using the gradient ascent pulse engineering  algorithm~\cite{GRAPE2005}.

\emph{\bfseries Realization of VQB.---}We implement our quantum circuit along with a post-processing procedure to achieve virtual quantum broadcasting of an arbitrary qubit state ($d=2$). Note that our approach can also be extended to the broadcasting of any finite-dimensional quantum system. The quantum circuit for broadcasting an arbitrary qubit state from qubit $B$ to qubit $A$ is shown in Fig.~\ref{Fig2}(a). The circuit involves three qubits: a probe qubit, qubit $A$, initialized in the maximally mixed state $\mathbb{I}/2$, and qubit $B$, which carries an arbitrary state $\rho$. After passing through the gates, a projective measurement is performed on the probe in the basis $\{ P_{0}, P_1\}$. Consequently, $AB$ collapses to $\{ \mathcal{B}^+(\rho),  \mathcal{B}^-(\rho) \},$
according to the probability distribution $\{ p_{0}=(d+1)/2d, p_1=(d-1)/2d\}$ (see SM~\cite{supp}). The desired outcome $\mathcal{B}(\rho)$ can then be obtained through classical post-processing as 
\begin{align}\label{Eq3}
\mathcal{B}(\rho)=d(p_0 \mathcal{B}^+(\rho) -p_1\mathcal{B}^-(\rho)).
\end{align}

We experimentally implemented VQB for various input states $\rho$, encompassing both pure and mixed states, with the results shown in Fig.~\ref{Fig2}(b-d). Since it is not feasible to perform projective measurements in an ensemble NMR system to directly obtain the collapsed states and probabilities, we reconstructed the quantum state of a three-bit quantum system~\cite{micadei2019reversing,oliveira2011nmr,PhysRevLett.129.100603,PhysRevLett.129.070502,fan2025solving,lin2024hardware} and applied mathematical post-processing to derive the experimental results for $p_0\mathcal{B}^+(\rho),$ and $p_1\mathcal{B}^-(\rho)$, and subsequently $\mathcal{B}(\rho)$~\cite{PhysRevLett.133.140602,xi2024experimental}.
The results for $2p_0\mathcal{B}^+(\rho), 2p_1\mathcal{B}^-(\rho)$, and $\mathcal{B}(\rho)$ corresponding to three input states, $\rho=\ket{0}\bra{0}$, $\ket{+}\bra{+}$ and $\mathbb{I}/4+\ket{+}\bra{+}/2$, are presented in Fig.~\ref{Fig2}(b-d), respectively, 
where $\ket{+}=(\ket{0}+\ket{1})/{\sqrt{2}}$.
The experimental results align closely with the theoretical predictions, demonstrating the success and reliability of our circuit and the post-processing procedure.

\begin{figure}[t!]
\centering
\includegraphics[width=1\linewidth]{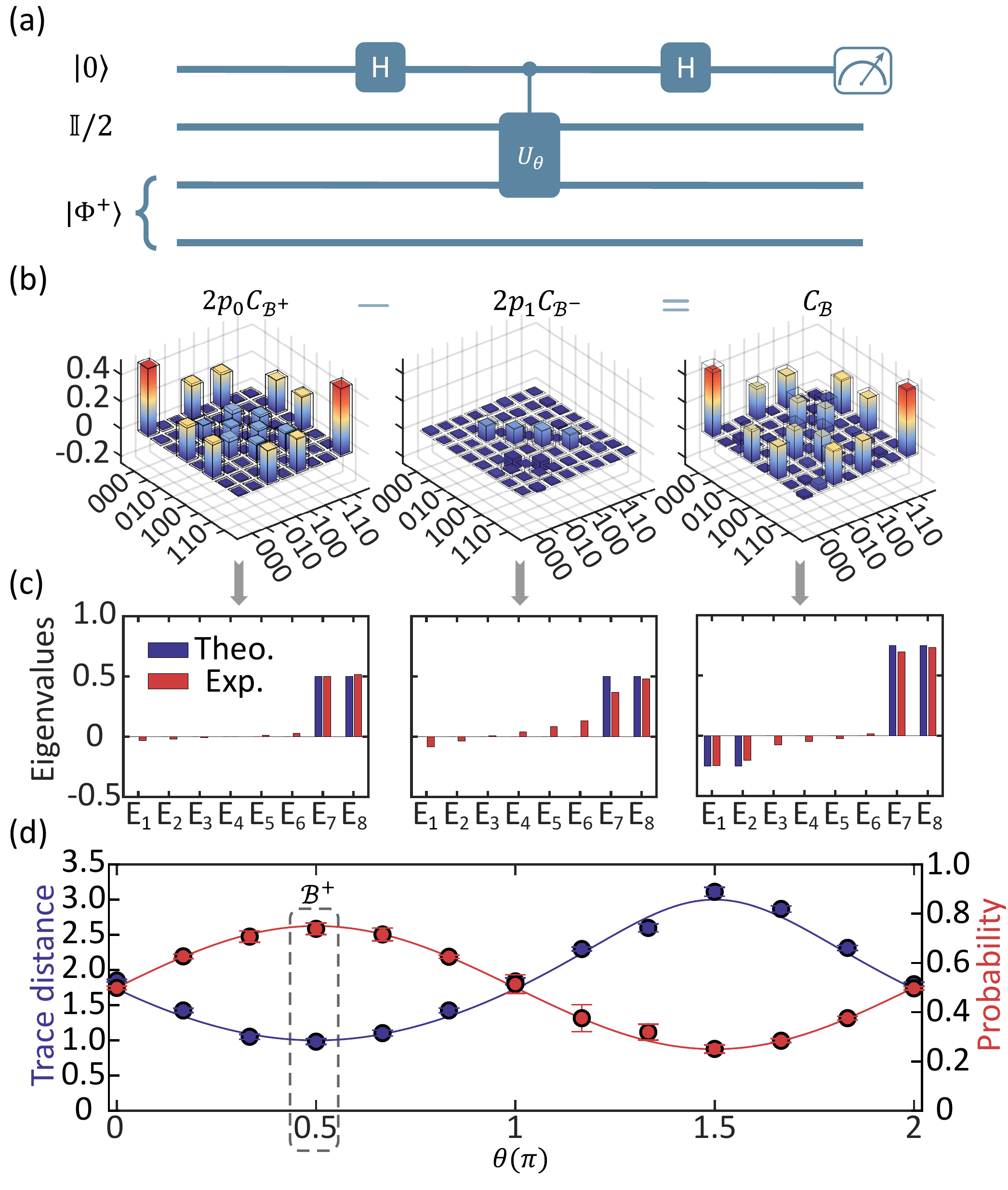}
\caption{ {\bf{Experimental validation of the universal cloner}}. (a) Quantum circuit for constructing the Choi states of the VQB map $\mathcal{B}$ and $\map{N}_\theta$. (b)  Experimental results for the Choi state of quantum broadcasting $C_{\mathcal{B}}$, expressed as a linear combination of $2p_0C_{\mathcal{B}^+}$ and $2p_1C_{\mathcal{B}^-}$. (c) Eigenvalues of the Choi states $C_{\mathcal{B}^+}$, $C_{\mathcal{B}^-}$  and $C_\mathcal{B}$. (d) The trace distance between $C_{\map B}$ and $C_{\map{N}_\theta}$ and the probability of obtaining $C_{\map{N}_{\theta}}$, as a function of $\theta$. The lowest trace distance is achieved at $\theta=\pi/2$ ($\map{N}_{\pi/2} = \map{B}^+$)  with the highest probability, thereby verifying that $\map{B}^+$ is the closest physical map to $\map B$. The experimental data represent the average of five measurements, and the error bars indicate the standard deviation across these repeated experiments~(see SM~\cite{supp}).}
\label{Fig3}
\end{figure}

\begin{figure*}[t!]
\centering
\includegraphics[width=1\textwidth]{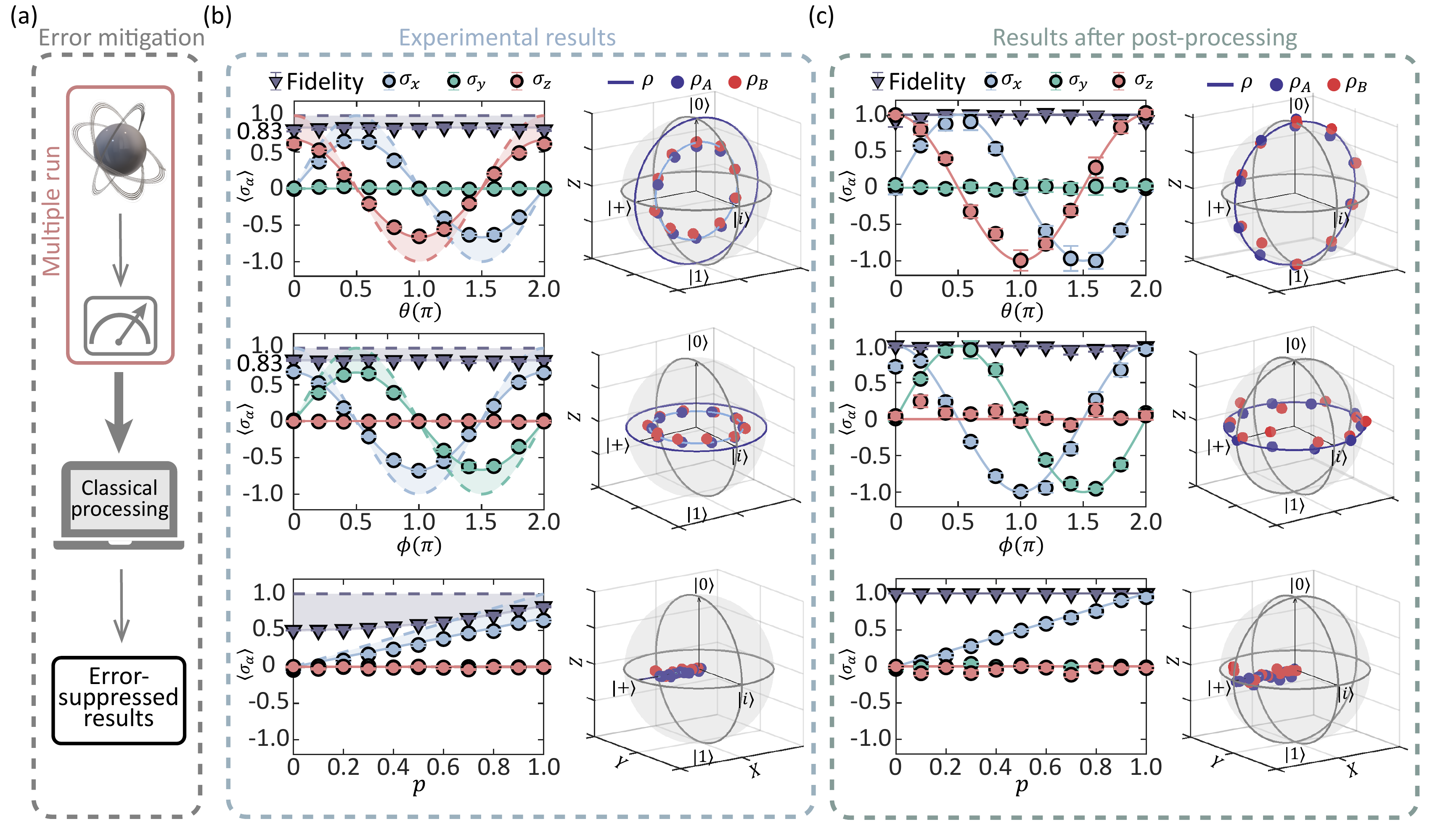}
\caption{{\bf{Experimental results of VQB following quantum error mitigation}}. (a) Illustration of the quantum error mitigation process. (b) Experimental results of the output of the universal cloner $\map{B}^+$.  From top to bottom, the expectation values of the Pauli operators on qubit $A$ are shown for the initial state $\rho$, representing three different categories of states: $\cos{(\theta/2)}\ket{0}+\sin{(\theta/2)}\ket{1}$, $\cos{(\pi/4)}\ket{0}+e^{i\phi}\sin{(\pi/4)}\ket{1}$, and $p\ket{+}\bra{+}+(1-p)\mathbb{I}/2$, with $\theta$ and $\phi$ varying from $0$ to $2\pi$, and polarization $p$ varying from $0$ to $1$, respectively. The experimentally constructed states $\rho_A$ and $\rho_B$ are represented by points on or inside the Bloch sphere. The fidelity $F(\rho_A, \rho) \approx 0.83$ for $\rho$ being pure is observed. The theoretical expectation values of Pauli operators on the input states $\rho$ are also shown as dashed lines in the right panels, with shaded areas representing the differences between the ideal quantum broadcasting and the universal cloner. 
We treat the shaded areas as errors in broadcasting quantum states, imposed by the constraints of quantum mechanics.
(c)  Results after post-processing. To mitigate the errors, we classically process the results of the universal cloner together with those of $\map{B}^-$, and present the results in (c). We observe that the fundamental error is significantly suppressed.  Each experimental data point represents the average of five trials, with error bars indicating the standard deviation of the measurements.}
\label{Fig4}
\end{figure*}

\emph{\bfseries  Validation of optimality of the universal cloner $\map{B}^+$.---}We design and implement an experiment to validate the universal cloner $\mathcal{B}^+$ as the closest CPTP map from a specially designed set of CPTP maps to the HPTP map $\map B$. The trace distance between two states is well understood, as it relates to the maximum probability of distinguishing between two quantum states. Inspired by this, distinguishing between two channels can be approached by transforming them into states and quantifying their corresponding distances. This leads to the diamond distance between two maps $\map{E}_1$ and $\map{E}_2 $ with the same input dimension $d$, given by $|| \map{E}_1-\map{E}_2||_\diamond= \max_{\sigma} \| \mathcal{I} \otimes\mathcal{E}_1 ( \sigma )- \mathcal{I} \otimes\mathcal{E}_2  (\sigma) \|_1
 $ where $\map{I}$ denotes the identity channel, $\sigma$ is a $2d-$dimensional system and $||\cdot||_1$ denotes the trace norm~\cite{wilde2011classical}. Coincidentally, the diamond distance for $\map{B}$ and $\map{B}^+$ is given when the output state $\sigma = |\Phi^+\rangle\langle \Phi^+| $ with  $|\Phi^+\rangle =\sum^{d-1}_{i=0} \ket{ii}/\sqrt{d}$~\cite{Parzygnat2024virtual}. Moreover, $
  (\map{I} \otimes \mathcal{E} )(|\Phi^+\rangle\langle \Phi^+|) =     C_{\mathcal{E}}$ is termed the Choi state of a map $\map E$~\cite{choi1975completely,jamiolkowski1972linear}. Therefore, the diamond distance between $\map{B}$ and $\map{B}^+$ equals to the trace distance between their Choi states. 
The experimental demonstration of the optimality of $\mathcal{B}^+$ is therefore based on realizing the Choi states of the maps of interest and evaluating their trace distance to the HPTP map $\map B$. 

We now describe a parameterized family  $\{\map{N}_{\theta}\}$ of CPTP maps, with $\map{B}^+$ as one of its elements, and the realization of the Choi state of  $\map{N}_\theta$ and $\map B$. 
Firstly, the design of the parameterized family $\{ \map{N}_\theta\}$ is inspired by the circuit in Fig.~\ref{Fig2}(a). Let $U(\theta)=ie^{-i\theta S}$ be the partial swap operator and $\theta\in[0, 2\pi]$ be a tunable parameter. By replacing the controlled-SWAP gate in Fig.~\ref{Fig2}(a) with a controlled-partial-SWAP gate, defined as
$\ket{0}\bra{0}\otimes\mathbb{I}+\ket{1}\bra{1}\otimes U(\theta)$,
and post-selecting the probe on $\ket{0}\bra{0}$, we obtain a parametrized family of CPTP maps on $\rho$, described by 
    $\map{N}_{\theta} (\rho)= \left(\mathbb{I}\otimes\mathbb{I}+ U(\theta) \right) (\rho \otimes \mathbb{I} ) \left(\mathbb{I}\otimes\mathbb{I}+ U^\dag(\theta) \right)/2(d+1)$.
A desired feature of the family $\{\map{N}_{\theta}\} $ is that it provides a smooth transition to the universal cloner $\map{B}^+$ since $\map{N}_{\pi/2} = \map{B}^+$. 
Secondly, according to the definition of the Choi state, the experimental circuit for realizing the Choi state of $\map{N}_\theta$ acting on a qubit is shown in Fig.~\ref{Fig3}(a). In this configuration, the input of the channel $\map{I} \otimes \map{N}_{\theta} $ is the Bell state $\ket{\Phi^+}= (\ket{00} + \ket{11})/\sqrt{2}.$
  Finally, the Choi state of $\map B$ can be constructed as follows: set $\theta= \pi/2$ and perform the projective measurement $\{P_0, P_1\}$ on the probe, which yields the collapsed states of the remaining three qubits, $C_{\map{B}^+}$ and $ C_{\map{B}^-} $, with probabilities $p_{0}$ and $p_1$, respectively (see SM~\cite{supp}). 
  The Choi state $C_{\map B}$ is constructed by  $d\left(p_0 C_{\map{B}^+}-p_1C_{\map{B}^-}\right)= C_{\map B}.$

 We experimentally construct the Choi states of $\map B$ as well as the series of CPTP maps $\{ \map{N}_\theta\}$, and subsequently calculate the trace distance between them.
 The experimental results and the corresponding theoretical predictions are presented in Fig.~\ref{Fig3}, demonstrating a high level of agreement between the experimental data and the theoretical predictions.
 In Fig.~\ref{Fig3}(b), we present the experimental results for the Choi states $2p_0 C_{\map{B}^+}$, $2p_1 C_{\map{B}^-}$, and $C_\mathcal{B}$. A linear map is CP if and only if its Choi state is positive~\cite{choi1975completely,jamiolkowski1972linear}. To verify the CP property, we show the eigenvalues of the Choi states $C_{\mathcal{B}^+}$, $C_{\mathcal{B}^-}$, and $C_{\mathcal{B}}$ in Fig.~\ref{Fig3}(c). The results show that $\mathcal{B}^\pm$ are CP maps, whereas the map $\mathcal{B}$ is not CP, reflecting the involvement of a post-processing operation. 
Finally, we vary $\theta$ from 0 to $2\pi$, project the probe onto $\ket{0}\bra{0}$ to record the Choi state $C_{\map{N}_{\theta}}$ of the maps $\map{N}_{\theta}$, and present the trace distance, $||C_{\map{B}} -C_{\map{N}_{\theta}}||_1$ as a function of $\theta$ in Fig.~\ref{Fig3}(d). 
The probability of obtaining the measurement outcome associated with $P_0$ as a function of $\theta$ is also presented. We observe that the minimal trace distance, with a value of $d-1 = 1$, is achieved at $\theta = \pi/2$ ($\map{N}_{\pi/2} = \map{B}^+$) with a maximal probability of $p_0 = 0.75$. This verifies that $\mathcal{B}^+$ is the closest physical process to the VQB map $\map{B}$ within our chosen CPTP maps.

\emph{\bfseries VQB as an error-mitigation protocol.---}The universal cloner $\map{B}^+$ represents the best physical protocol for broadcasting quantum states~\cite{buifmmode1996quantum,werner1998optimal,bruss1198optimal,bruss1998optimalprl,chiribella2010quantum,harrow2013church}, though its outcome does not achieve perfect cloning fidelity. Here, we treat this imperfection—specifically, the deviation of the expectation values of Pauli operators in quantum state tomography—as error. Our circuit not only implements $\map{B}^+$ but also the universal antisymmetrizer $\map{B}^-$. We demonstrate how to mitigate this error in the outcome of $\map{B}^+$ through post-processing with $\map{B}^-$, thereby highlighting VQB as an effective error mitigation protocol for quantum broadcasting.

The overall error mitigation scheme is presented in Fig.~\ref{Fig4}(a). We implement the quantum circuit depicted in Fig.~\ref{Fig2}(a) with different input states $\rho$. We perform the measurement $\{P_{0}, P_1\}$ on the probe qubit and then implement the Pauli observables $\sigma_\alpha, \alpha= x, y, z,$ on the two qubits $A$ and $B$. In Fig.~\ref{Fig4}(b), we present the expectation values $\langle \sigma_\alpha \rangle_A=\Tr (\sigma_\alpha \rho_A)$ in the case of $P_0$ being measured (such that $\rho_{AB}=\map{B}^+(\rho)$).  Using all the expectation values $\langle \sigma_\alpha \rangle$ for both $A$ and $B$, the experimentally constructed quantum states $\rho_A$ and $\rho_B$ are presented as points on or inside the Bloch sphere alongside the expectation values. Moreover, fidelity $F(\rho_A, \rho) \approx 0.83$ is also shown when $\rho$ is pure. Our experimental results are in good agreement with the theoretical predictions, yet we observe a significant gap between the experimental results and those of the input state $\rho$, represented by the shaded areas. We refer to these areas as fundamental errors imposed by quantum mechanics.
We utilize the experimental results of the antisymmetrizer $\map{B}^-$ to mitigate this error. We classically process the data according to Eq.~\eqref{Eq3}, and present the corresponding results in Fig.~\ref{Fig4}(c).
We observe that the fundamental errors are significantly suppressed, and the fidelities are near 1, demonstrating the effectiveness of VQB as an error mitigation protocol for broadcasting arbitrary quantum states.

\emph{\bfseries Conclusions.---} We have reported the first experimental implementation of VQB, designing and realizing a protocol on an NMR system to virtually broadcast arbitrary qubit states. Our experimental results confirmed that the universal cloner is the closest physical map to VQB. Furthermore, we demonstrated that VQB can be interpreted as a protocol for mitigating `fundamental errors' imposed by quantum mechanics.  An alternative approach to realizing the VQB can be found in Ref.~\cite{buscemi2013direct}.

The proposed protocol builds on the probabilistic realization of the universal cloner and the universal antisymmetrizer within the same circuit. By allowing post-processing of the data, our protocol can, in principle, achieve perfect cloning fidelity, thereby enabling quantum broadcasting for arbitrary quantum states of any dimension. In addition, our approach is efficient, as no data is wasted given the probabilistic nature of our method. Other methods for achieving better cloning (e.g., than the universal cloner) have specific state requirements. For instance, Refs.~\cite{PhysRevLett.80.4999,chen2011exp} requires the states to be linearly independent, Ref.~\cite{bruss1198optimal} necessitates prior knowledge of the state, and Ref.~\cite{buifmmode1996quantum} applies specifically to quantum states lying on the equator of the Bloch sphere.

Our work opens avenues for further exploration and applications. 
In our experiment, all qubits—except for the probing qubit—remain untouched, allowing them to be reused in subsequent experiments, particularly for tasks requiring multiple copies. For example, our approach can facilitate the broadcasting of an initial state that is hard to prepare~\cite{tang2021quantum}. When combined with classical communication, it also enables state broadcasting to distant parties for distributed quantum computing and sensing, etc. Moreover, our protocol may be useful in virtual quantum cooling~\cite{PhysRevX.9.031013}, where multiple copies of a thermal state at a higher physical temperature are used to simulate the cooling of the thermal state.

\emph{\bfseries Acknowledgment.---} We thank James Fullwood for suggesting this experiment and Zhenhuan Liu for his feedback.
This work is supported by Pearl River Talent Recruitment Program (2019QN01X298), Guangdong Provincial Quantum Science Strategic Initiative (GDZX2303001, GDZX2200001, GDZX2403004), and Guangdong Basic and Applied Basic Research Foundation (2025A1515011599). Y. L. is supported by China Scholarship Council (No.202408060137). X. L. is supported by the National Research Foundation, Prime Minister’s Office, Singapore under its Campus for Research Excellence and Technological Enterprise (CREATE) programme.

\bibliography{references.bib}

\appendix

\onecolumngrid
\newpage

\begin{center}
	\textbf{\large Supplementary Material for Experimental Virtual Quantum Broadcasting}\\
	 \vspace{2ex}
	\text{Yuxuan Zheng$^1$, Xinfang Nie$^{2,1}$, Hongfeng Liu$^1$, Yutong Luo$^{3,4}$, Dawei Lu$^{1,2}$, Xiangjing Liu$^{5,6,7}$ }\\
		 \vspace{1ex}
\textit{$^1$ Department of Physics, Southern University of Science and Technology, Shenzhen 518055, China} \\
\textit{ $^2$ Quantum Science Center of Guangdong-Hong Kong-Macao Greater Bay Area, Shenzhen 518045, China}\\
\textit{$^3$ School of Physics, Trinity College Dublin, College Green, Dublin 2, D02 K8N4, Ireland}\\
\textit{$^4$ Trinity Quantum Alliance, Unit 16, Trinity Technology and Enterprise Centre, Pearse Street, Dublin 2, D02 YN67, Ireland}\\
\textit{ $^5$ CNRS@CREATE, 1 Create Way, 08-01 Create Tower, Singapore 138602, Singapore}\\
\textit{ $^6$ MajuLab, CNRS-UCA-SU-NUS-NTU International Joint Research Unit, Singapore}\\
\textit{$^7$ Centre for Quantum Technologies, National University of Singapore, Singapore 117543, Singapore}\\
\end{center}

\section{A. Quantum spatiotemporal formalism associated with virtual quantum broadcasting}

This section provides an introduction to the quantum spatiotemporal formalism associated with the virtual quantum broadcasting (VQB) map. The quantum spatiotemporal formalism is known as the pseudo-density matrix formalism~\cite{fitzsimons2015quantum,liu2023quantum}. The pseudo-density matrix represents a type of quantum state over time, where states are defined across both space and time.  We will briefly introduce the formalism and then show how it relates to virtual quantum broadcasting.

The pseudo-density matrix (PDM) formalism generalizes the standard density matrix by assigning a separate Hilbert space to each moment in time~\cite{fitzsimons2015quantum,liu2023quantum}. Let us consider assigning Hilbert spaces to the following two moments in time: a quantum system $\rho$ consists of $n$ qubits at time $t_A$
 undergoing a quantum channel 
$\mathcal{N}$, and arriving at time $t_B$. The two-time PDM $R_{AB}$  is envisioned as $n$-qubit Pauli measurements performed at the two times $t_A$ and $t_B$ and it is  defined as follows:
\begin{align}\label{eq: DefPDM}
& R_{AB} = \frac{1}{4^{n}} \sum^{4^n-1}_{i_A,i_B =0 }  \langle \{ \sigma_{i_A}, \sigma_{i_B} \} \rangle  \sigma_{i_A} \otimes  \sigma_{i_B} .
\end{align}
In this expression, $\sigma_{i_\alpha} \in \{ \mathbb{I}, \sigma_x, \sigma_y, \sigma_z \} ^ {\otimes n}$  represents an $n$-qubit Pauli matrix at time $t_\alpha$
  and $\langle \{\sigma_{i_A}, \sigma_{i_B} \}\rangle$ denotes the expectation value of the product of the outcomes of the two measurements. The PDM has several nice properties. The PDM is Hermitian and has a unit trace, it can possess negative eigenvalues, which, while not necessary, serve as a sufficient indicator of quantum temporal correlations.  Taking the partial trace of a PDM still results in a valid PDM. Specifically, when left with only one time, the reduced PDM give a valid density matrix, i.e.,
  \begin{align}
      \rho_A &= \Tr_B R_{AB} =\rho, \nonumber\\
      \rho_B &= \Tr_A R_{AB} = \mathcal{N}(\rho).
  \end{align}
  Interestingly, 
if the measurement scheme for the observable $\sigma_i$ at each time is set to be the projectors that project the state onto the $\pm 1$ eigenspaces of $\sigma_i$, i.e., 
$$
\left\{P^i_+ = \frac{\mathbb{I} + \sigma_i}{2} , P^i_- = \frac{\mathbb{I} - \sigma_i}{2}  \right\},
$$
 the 2-time PDM has the closed-form expression~\cite{liu2023quantum}
\begin{align}
\label{eq: closedPDM}
R_{AB}= \frac{1}{2} \{ \rho_A\otimes \mathbb{I}_B,  \, M_{\mathcal{N}}  \},
\end{align}
where  $\{ ,\}$ denotes the anticommutator, $\rho_A$ denotes the quantum state at time $t_A$ and $M_{\mathcal{N}}$ is a variant of the Choi-Jamilkowski (CJ) matrix of the channel $\mathcal{N}$, which is defined as~\cite{choi1975completely,jamiolkowski1972linear}
\begin{align}
    M_{\mathcal{N}}=& \sum_{i,j} \mathcal{I}\otimes \mathcal{N } (\ket{ii}\bra{jj})^{T} 
    = \sum_{ij}\ket{i}\bra{j} \otimes \mathcal{N}(\ket{j}\bra{i}),
\end{align}
where $T$ denotes the partial trace.  Although the above results were derived in qubit systems, the closed form of the PDM has been generalized to arbitrary 
$d$-dimensions~\cite{fullwood2024operator}. It is worth noting that the PDM of the form given in Eq.~\ref{eq: closedPDM} is sometimes also referred to as the quantum state over time~\cite{fullwood2022on,Parzygnat2024virtual}.    

We are now in a good position to state the operational meaning of the output of VQB. As shown in Fig.~\ref{Fig1}, when the intermediate channel $\mathcal{N}$ is the identity channel $\mathcal{I}$, the corresponding CJ matrix $M_{\mathcal{I}}=S$ where $S$ denotes the swap operator. The corresponding PDM is given by
\begin{align}
    R_{AB}= \frac{1}{2} \{\rho 
    \otimes \mathbb{I}, S \}, \ \ \text{with} \ \       \Tr_B R_{AB} =\rho= \Tr_A R_{AB}.
\end{align}
Define the map $\mathcal{B}$ by its action on $\rho$ 
$$ \mathcal{B}(\rho) = \frac{1}{2} \{\rho \otimes \mathbb{I}, S \}.$$  
This map is exactly the virtual broadcasting map.  In the scenario of qubit systems, it can therefore be interpreted as a quantum system, $\rho$, at time $t_A$, undergoing an identity channel and arriving at time $t_B$, with $n$-qubit Pauli measurements performed at both times $t_A$ and $t_B$. The interpretation also applies to quantum systems of arbitrary dimension $d$ by replacing the Pauli measurements with the so-called light-touch observables~\cite{fullwood2024operator}.

\section{B. Procedure for simulating virtual quantum broadcasting} 

This section outlines a detailed procedure for simulating the HPTP map associated with virtual quantum broadcasting. The procedure involves two steps: 1) quantum circuit design and 2) post-selection and classical processing.

 We begin with an overview of our approach. 
Denoting the two CPTP maps as
\begin{align}\label{eq:Bpm}
    \mathcal{B}^+ (\rho) = \frac{2}{d+1} \Pi^{+} (\mathbb{I} \otimes \rho) \Pi^{+} \ \text{and }\ \mathcal{B}^- (\rho) = \frac{2}{d-1} \Pi^{-} (\mathbb{I} \otimes \rho) \Pi^{-},
\end{align}
where $\Pi^{\pm} = \frac{1}{2}(\mathcal{I} \otimes \mathcal{I} \pm S) $ and $d$ is the dimension of the quantum system.
As suggested in Ref.~\cite{Parzygnat2024virtual}, the HPTP map $\mathcal{B}$ associated with quantum virtual broadcasting can be decomposed into 
\begin{align}\label{eq:Bdecomposed}
    \mathcal{B}= \frac{d+1}{2}\mathcal{B}^+ - \frac{d-1}{2}\mathcal{B}^-.
\end{align}
Since the two operators $\mathcal{I} \otimes \mathcal{I} $ and $S$ are unitaries, the two maps $\mathcal{B}^{\pm}$ can be derived using the technique of the linear combination of unitaries (LCU)~\cite{childs2012hamiltonian}  and the outcome of the HPTP map $\mathcal{B}$ can be obtained by classically processing the outcomes of $\mathcal{B}^\pm$.

\begin{figure}[hbt!]
\centering
\includegraphics[width=0.5\linewidth]{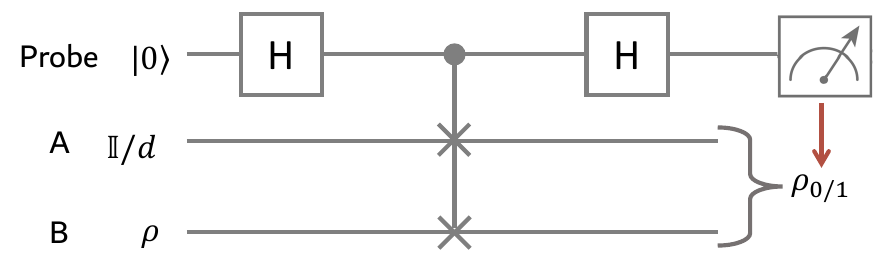}
\caption{ {\bf{Circuit for  virtual quantum broadcasting.}} }
\label{fig:s1}
\end{figure}

\emph{\bfseries Step 1: Quantum circuit design.} Our circuit is demonstrated in Fig.~\ref{fig:s1}. The design of the circuit takes inspiration from the linear combination of unitaries~\cite{childs2012hamiltonian}. The circuit involves three systems: a quantum system $A$ we want to broadcast, a quantum system $B$ we want to broadcast to, and an ancilla control system $C$. The three systems are initialized in 
\begin{align}
    \rho_{in} = \ket{0}_C\bra{0} \otimes \frac{\mathbb{I}_B}{d} \otimes \rho_A.
\end{align}
They undergo the following transformation before the measurement:
\begin{align}\label{eq:output}
|0\rangle \langle 0|  \otimes \frac{ \mathbb{I}}{d} \otimes \rho 
\xrightarrow{\text{Hadamard}}
&|+\rangle \langle +| \otimes \frac{ \mathbb{I}}{d} \otimes \rho  \nonumber\\
\xrightarrow{\text{C-SWAP}} &\frac{1}{2} \left( |0\rangle \langle 0| \otimes \frac{\mathbb{I}}{d} \otimes \rho +
|1\rangle \langle 1| \otimes \rho \otimes \frac{\mathbb{I}}{d} +
|0\rangle \langle 1| \otimes (\frac{\mathbb{I}}{d} \otimes \rho)S +
|1\rangle \langle 0| \otimes S(\frac{\mathbb{I}}{d} \otimes \rho)
\right) \nonumber\\
\xrightarrow{\text{Hadamard}}
&\frac{1}{2} \left( |+\rangle \langle +| \otimes \frac{\mathbb{I}}{d} \otimes \rho +
|-\rangle \langle -| \otimes \rho \otimes \frac{\mathbb{I}}{d} +
|+\rangle \langle -| \otimes (\frac{\mathbb{I}}{d} \otimes \rho)S +
|-\rangle \langle +| \otimes S(\frac{\mathbb{I}}{d} \otimes \rho)
\right) \nonumber\\
=& \rho_{out}
\end{align}

\emph{\bfseries Step 2: Post-selection and classical processing.} In this step, the control system $C$ is measured in the computational basis. Depending on the measurement outcome,  the corresponding states of the compound system $AB$ are then recorded and used to construct the outcome of the VQB map.

The detailed construction is as follows. If the control system is projected on $\ket{0}\bra{0}$, the conditional state $\rho_0$ of $AB$ is 
\begin{align}
p_0\rho_{0}
&=  \frac{\mathbb{I}\otimes\rho+\rho\otimes \mathbb{I}+ (\mathbb{I}\otimes\rho) S+ S(\mathbb{I}\otimes \rho)}{4d }  \nonumber\\
&= \frac{1}{d} \Pi^+ (\mathbb{I}\otimes \rho ) \Pi^+  \nonumber\\
&= \frac{d+1}{2d} \mathcal{B}^+(\rho). 
\end{align}
If the control system is projected on $\ket{1}\bra{1}$, the conditional state $\rho_1$ of $AB$ is 
\begin{align}
p_1\rho_{1}
=&   \frac{\mathbb{I}\otimes\rho+\rho\otimes \mathbb{I}- (\mathbb{I}\otimes\rho) S- S(\mathbb{I}\otimes \rho)}{4d }  \nonumber\\
=& \frac{1}{d} \Pi^- (\mathbb{I} \otimes \rho) \Pi^- \nonumber\\
= &\frac{d-1}{2d}  \mathcal{B}^-(\rho) . 
\end{align}
The above calculation demonstrates that the postselected states of $AB$ can indeed be viewed as arising from the LCU.
The probabilities are obtained via
\begin{align}\label{eq: Cprob}
p_{0}&=\frac{1}{4d}\Tr\left[{ \mathbb{I}\otimes\rho+\rho\otimes \mathbb{I}+ 2\mathcal{B}(\rho)}\right] = \frac{d+1}{2d}, \nonumber\\
p_{1}&=\frac{1}{4d}\Tr\left[{ \mathbb{I}\otimes\rho+\rho\otimes \mathbb{I}- 2\mathcal{B}(\rho)}\right] = \frac{d-1}{2d},
\end{align}
where $\mathcal{B}(\rho)=\frac{1}{2}\{\rho\otimes \mathbb{I},S\}$ and $\Tr \mathcal{B}(\rho)=1$ are used. From Eq.~\eqref{eq:Bdecomposed}, the outcome of the VQB map is constructed via
\begin{align}
    \mathcal{B}(\rho) &= \frac{d+1}{2}\mathcal{B}^+(\rho) - \frac{d-1}{2}\mathcal{B}^-(\rho) \nonumber\\
    &= d (p_0\rho_0- p_1\rho_1).
\end{align}

We conclude that the HPTP map $\mathcal{B}$, associated with virtual quantum broadcasting, can be simulated by a CPTP map (our quantum circuit) followed by post-selection and classical processing.

\section{C. Quantum circuit for preparing the Choi state of the VQB map and the optimal quantum cloning map }
\begin{figure}[hbt!]
\centering
\includegraphics[width=0.5\linewidth]{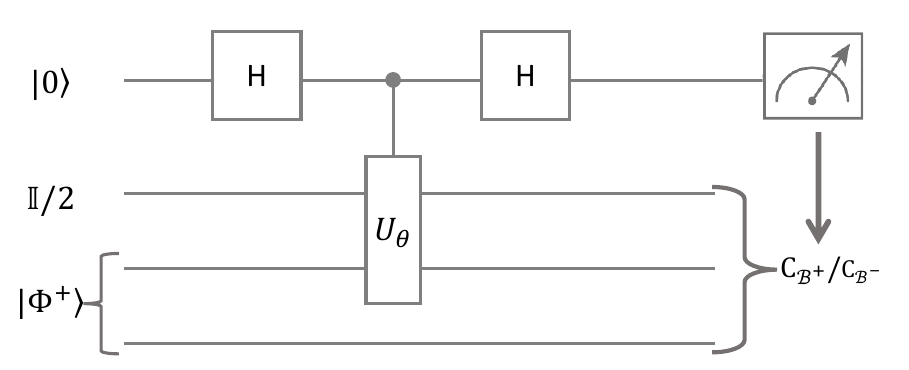}
\caption{ Circuit for  preparing the Choi state of the VQB map and the optimal quantum cloning map.}
\label{fig:s2}
\end{figure}

This section discusses the method for constructing the Choi state of the VQB map $\map B$ and the optimal quantum cloning map $\map{B}^+$.
These two Choi states will be used to examine the CP property and compute the distance between the two maps.

Firstly, we consider the construction of the universal cloner $\map{B}^+$. The Choi state for any linear map $\map N$ acting on a quantum system with dimension $d$ is defined as~\cite{choi1975completely,jamiolkowski1972linear} 
\begin{align}
    C_{\mathcal{N}}=\map{I} \otimes \map{N}   ( \ket{\Phi^+} \bra{\Phi^+})
     = \frac{1}{d} \sum^{d-1}_{i,j=0} \mathcal{I}\otimes \mathcal{N } (\ket{ii}\bra{jj}) 
    = \frac{1}{d}\sum_{ij}\ket{i}\bra{j} \otimes \mathcal{N}(\ket{i}\bra{j}),
\end{align}
where $\ket{\Phi^+} = \sum^{d-1}_{i=0} \ket{i,i}/ \sqrt{d} $ denotes the maximally entangled state.
The Choi state of the map $\map{B}^+$ is therefore calculated by
\begin{align}\label{eq:choiB+}
     C_{\map{B}^+} 
    =& \frac{1}{d}\sum_{i,j} |i\rangle\langle j|\otimes \map{B}^+(\ket{i}\bra{j}) 
    =   \frac{2}{d(d+1)}  \sum_{i,j} \ket{i}\bra{j} \otimes  \Pi^{+} (\mathbb{I} \otimes \ket{i}\bra{j}) \Pi^{+}
\end{align}
It is straightforward to that the Choi matrix $C(\hat{B}^+)$ is positive. This expression also suggests a similar way of preparing the Choi state of $\map{B}^+$ as the map $\map{B}^+$ itself. The circuit is given in Fig.~\ref{fig:s2}

The Choi state of $\map B$ can be prepared via classical processing. Similarly one can calculate the Choi state of the map $\map{B}^-$ by
\begin{align}\label{eq:choiB-}
     C_{\map{B}^-} 
    = \frac{1}{d}\sum_{i,j} |i\rangle\langle j|\otimes \map{B}^-(\ket{i}\bra{j}) =   \frac{2}{d(d-1)}  \sum_{i,j} \ket{i}\bra{j} \otimes  \Pi^{-} (\mathbb{I} \otimes \ket{i}\bra{j}) \Pi^{-}.
\end{align}
Combine Eq.~\eqref{eq:choiB+} and Eq.~\eqref{eq:choiB-}, the Choi state of $\map B$ is given by 
\begin{align}
     C_{\map{B}} 
    =& \frac{1}{d}\sum_{i,j} |i\rangle\langle j|\otimes \map{B}(\ket{i}\bra{j}) \nonumber\\
    =&   \frac{1}{d}  \sum_{i,j} \ket{i}\bra{j} \otimes  \left(\frac{d+1}{2}\mathcal{B}^+(\ket{i}\bra{j}) - \frac{d-1}{2}\mathcal{B}^-(\ket{i}\bra{j}) \right) \nonumber\\
    =& d (p_0 C_{\map{B}^+} - p_1 C_{\map{B}^-} ),
\end{align}
where the probabilities are given in Eq.~\eqref{eq: Cprob}.

\section{D. Experimental Details}
Our experiments were carried out on a nuclear magnetic resonance (NMR) quantum processor, where nuclear spins within a molecule are used to encode qubits. We begin by providing a comprehensive characterization of the NMR system, detailing key aspects such as sample preparation, control methods, and measurement protocols. Next, we outline the procedure employed to generate a pseudo-pure state (PPS).

\begin{figure}[htbp]
\centering
\includegraphics[width=0.9\linewidth]{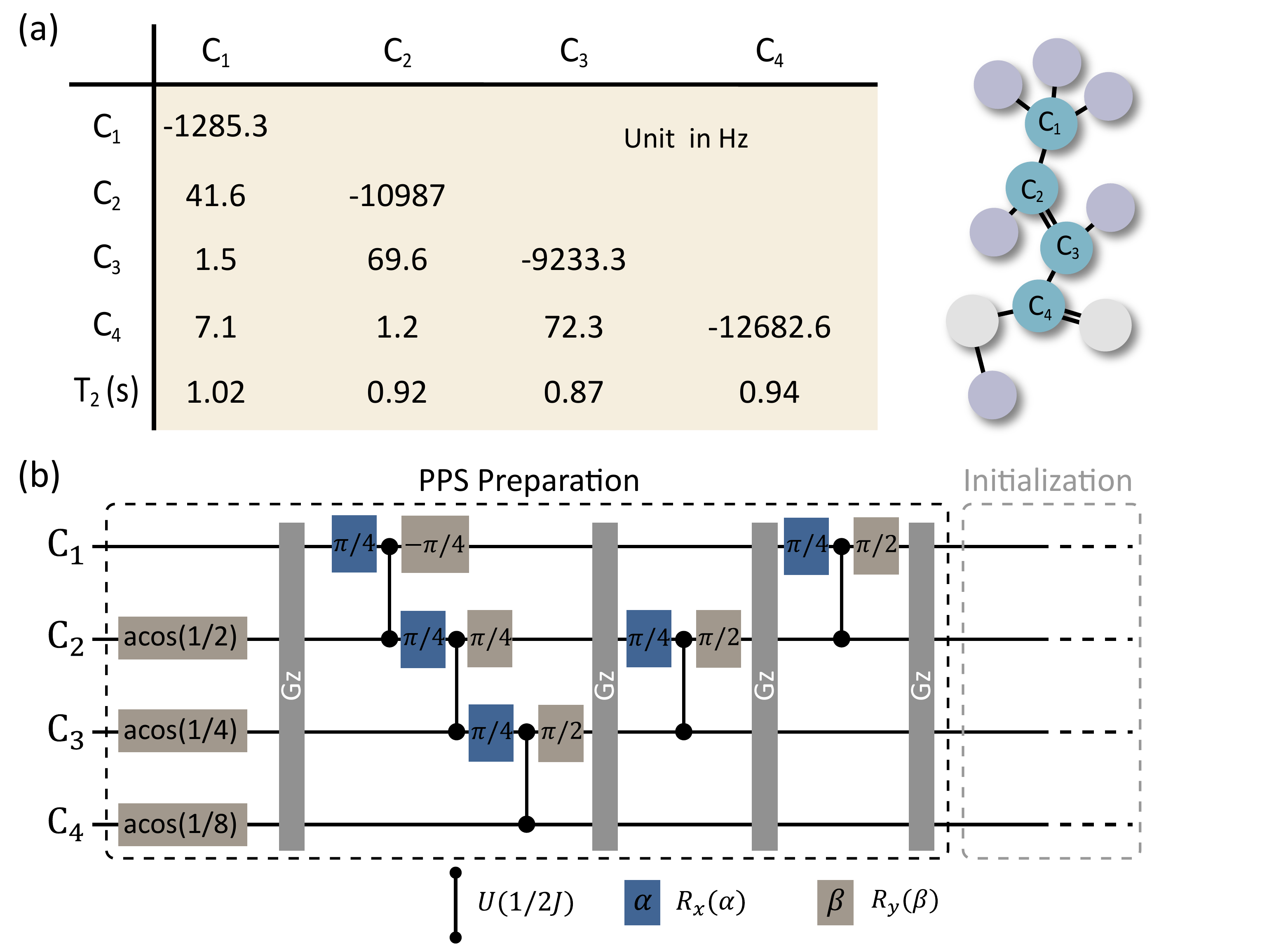}
\caption{(a) The NMR parameters and molecular structure of $^{13}$C-labeled trans-crotonic acid. In the table, the chemical shifts and the scalar coupling constants (in Hz) are listed by the diagonal and off-diagonal numbers, respectively. The relaxation time $T_2$ (in seconds) are shown at the table. (b) NMR pulse sequence to prepare the 4-qubit system to the PPS. The Purple and green rectangles indicate the $R_x$ and $R_y$ rotation gate, respectively. The grey rectangles mean the Gz pulse. The gradient-field pulse, denoted by Gz, are used to eliminate all coherence from the instantaneous state.}
\label{molecular structrue&pps}
\end{figure}

\textit{Characterization.}--In this experiment, a 4-qubit quantum processor was implemented using $^{13}$C-labeled trans-crotonic acid dissolved in $d_6$ acetone. The molecular structure and corresponding parameters are shown in Fig.~\ref{molecular structrue&pps}(a). Qubits Q1 through Q4 correspond to $^{13}$C$_1$ through $^{13}$C$_4$, respectively. The methyl group (gray balls) and all hydrogen atoms (purple balls) were decoupled throughout the experiment. The total Hamiltonian $\mathcal{H}_\text{tot}$ for the system includes both the internal Hamiltonian $\mathcal{H}_\text{int}$ and the control Hamiltonian $\mathcal{H}_\text{con}$, expressed as:  
\begin{align}
{\mathcal{H}_{{\rm{tot}}}} =&{\mathcal{H}_{{\rm{int}}}} +{\mathcal{H}_{{\rm{con}}}} \\\nonumber
 =&\sum\limits_{i = 1}^4 {\pi {\nu_i}\sigma _z^i} + \sum\limits_{1 \le i < j \le 4}^4 {\frac{\pi }{2}{J_{ij}}\sigma _z^i} \sigma _z^j\\\nonumber
 &-B_1\sum\limits_{i = 1}^4 \gamma_i[\cos(\omega_{rf}t+ \phi)\sigma_x^i+\sin(\omega_{rf}t+ \phi)\sigma_y^i],
\end{align}
where $\nu_i$ represents the chemical shift of the $i$-th spin, and $J_{ij}$ is the scalar coupling constant between the $i$-th and $j$-th nuclei. The parameters $B_1$, $\omega_{rf}$, and $\phi$ correspond to the amplitude, frequency, and phase of the control pulses, respectively.  

\textit{Pseudo-pure state preparation.}--At room temperature, the thermal equilibrium state of the 4-qubit NMR system is a highly mixed state, described by  
\begin{equation}
\rho_{\text{eq}} = \frac{\mathbb{I}}{16} + \epsilon\sum^4_{i=1}\sigma_z^i,
\label{eqstate}
\end{equation}
where $\mathbb{I}$ is the $16 \times 16$ identity matrix, and $\epsilon$ represents the polarization, approximately $10^{-5}$. This state is unsuitable as an initial state for quantum computing. Various initialization techniques are available, including the spatial averaging method, line-selective transition method, time averaging method, and cat-state method. In our experiment, we employed the spatial averaging method to initialize the NMR system, using the pulse sequence shown in Fig.~\ref{molecular structrue&pps}(b). In the circuit diagram, colored rectangles represent single-qubit rotations implemented with rf pulses, while two-qubit gates are realized through scalar coupling between spins combined with shaped pulses. This pulse sequence transforms the equilibrium state in Eq.~\eqref{eqstate} into a pseudo-pure state (PPS), expressed as:  
\begin{equation}
\rho_{\text{PPS}} = \frac{1-\epsilon'}{16}\mathbb{I} + \epsilon'\op{0000}{0000}.
\label{eq:PPS}
\end{equation}
The dominant component, the identity matrix $\mathbb{I}$, remains unchanged under unitary transformations and is undetectable in NMR experiments. Consequently, the quantum system can effectively be treated as the pure state $\op{0000}{0000}$, despite its actual mixed nature.  
In our experimental setup, each segment of the quantum circuit was combined into one unitary operation, separated by three gradient pulses. The corresponding rf pulses were then optimized using an optimal-control algorithm. The shaped pulses used in the experiments had durations of 3 ms, 20 ms, 15 ms, and 15 ms, respectively, with fidelities exceeding 99.5$\%$.

\begin{figure}[htbp]
\centering
\includegraphics[width=1\linewidth]{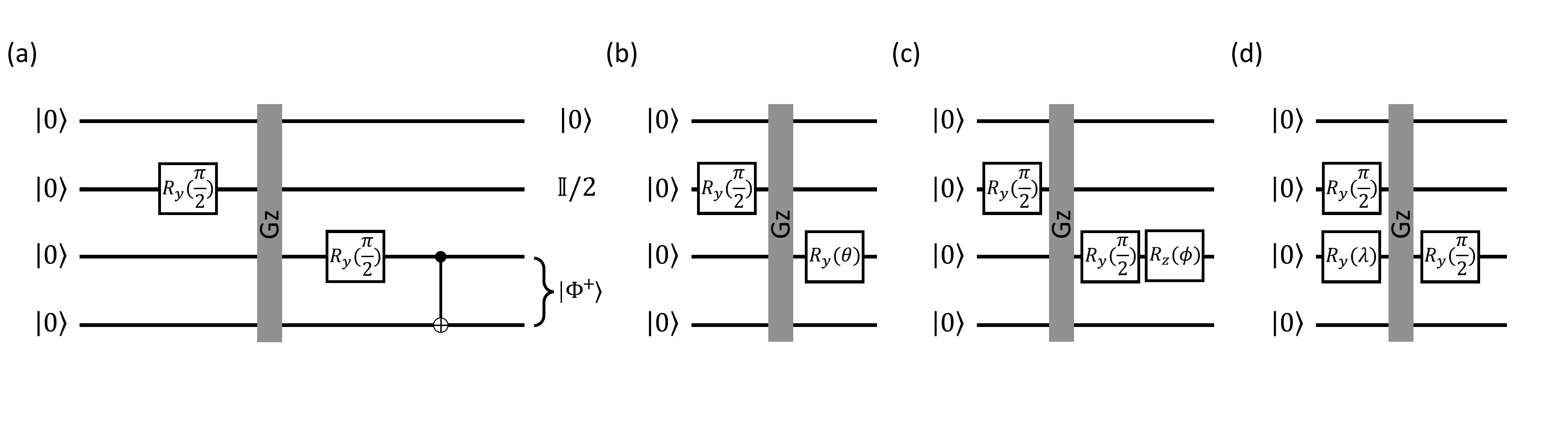}
\caption{(a) Initialization circuit for the universal cloner experiment. (b-d) Initialization circuits for preparing: (b) pure states of the form $\cos{(\theta/2)}\ket{0}+\sin{(\theta/2)}\ket{1}$, (c) pure states of the form $\cos{(\pi/4)}\ket{0}+e^{i\phi}\sin{(\pi/4)}\ket{1}$, and (d) mixed states of the form $p\ket{+}\bra{+}+(1-p)\mathbb{I}/2$, respectively.}
\label{Initialization}
\end{figure}

\textit{Initialization.}--Following the preparation of the PPS, the next step is initialization. For the validation of the universal cloner experiment, the initial circuit is shown in Fig.~\ref{Initialization}(a). Qubit Q2 is prepared in a maximally mixed state by applying a rotation operation $R_y(\frac{\pi}{2})$ followed by a gradient-field pulse Gz. Meanwhile, Qubits Q3 and Q4 are entangled into the state $\ket{\Phi^+}$ using a combination of rotation operations $R_y(\frac{\pi}{2})$ and CNOT gates. Figures~\ref{Initialization}(b-d) illustrate the initialization of all initial states used in the experiments. 
For the initialization shown in Fig.~\ref{Initialization}(d), the mixed state $\rho = p\ket{+}\bra{+} + (1-p)\mathbb{I}/2$ is prepared using two single-qubit gates and a gradient-field pulse. The first single-qubit rotation, $R_y(\lambda)$, prepares the qubit in the state $\sqrt{(1+p)/2}\vert0\rangle + \sqrt{(1-p)/2}\vert1\rangle$, where $\alpha = \text{arccos}\sqrt{\langle \sigma_x \rangle^2 + \langle \sigma_y \rangle^2 + \langle \sigma_z \rangle^2}$ and $\langle \sigma_i \rangle = \text{Tr}(\rho_A \sigma_i)$. A gradient-field pulse is then applied to dephase the state, removing coherence and transforming it into the mixed state $(1-p)\mathbb{I}/2 + p\vert0\rangle\langle0\vert$, ensuring that it matches the purity of the target state $\rho_A$.
Finally, a second single-qubit rotation, $R_y(\pi/2)$, is used to evolve the state into $\rho_A$. Simultaneously, a Hadamard gate is applied to the probe qubit, preparing it in the state $\vert+\rangle\langle+\vert$.

\textit{Quantum state tomography.}--
In an NMR quantum processor, the experimental sample consists of a large number of identical molecules rather than a single molecule. As a result, the measurements performed by the NMR system represent ensemble averages over these molecules. After an operation, the nuclear spins precess around the $B_0$ direction and gradually relax back to thermal equilibrium. This precession induces an electrical signal in the $x,y$-plane, allowing the NMR system to measure only the transverse magnetization components, specifically the expectation values of $\sigma_x$ and $\sigma_y$.
In a four-qubit NMR quantum processor, the signal from each nuclear spin is typically split into eight peaks due to couplings between different nuclei. Based on spin dynamics in NMR, each peak's signal comprises both real and imaginary components, encoding the expectation values of the Pauli matrices $\sigma_x$ and $\sigma_y$ for the observed spin, respectively. Consequently, the NMR system can measure the expectation values of single-quantum coherence operators that involve $\sigma_x$ or $\sigma_y$ for the target qubit, and $\sigma_z$ or $I$ for the remaining qubits.
For example, the readout pulse sequence $R^1_y(-\frac{\pi}{2})+R^3_y(-\frac{\pi}{2})$ can be used to measure the observable $\sigma_xI\sigma_x\sigma_x$ by transferring it to $\sigma_zI\sigma_z\sigma_x$.

A quantum state can be decomposed in the complete Pauli bsis $\prod^n_{i=1}\otimes \sigma^i_{0,x,y,z}$ with the Pauli matrices $\sigma_0=I,\sigma_x,\sigma_y,$ and $\sigma_z$. The expectation values of each Pauli basis can be measured in the experiments.

\section{E. Experimental Error Analysis}
In NMR experiments, we deal with an ensemble of approximately $10^{17}$ molecules, which inherently stabilize measurements due to the central limit theorem.  
Our pulse sequence lasts only about 30 ms, significantly shorter than the $T_2$ relaxation time (on the order of seconds), minimizing signal loss from decoherence.
Moreover, the shaped pulse used for the unitary operator has a simulated fidelity exceeding 0.995, ensuring precise coherent control. 
Taken together, these factors keep the overall experimental error small.Each measurement is repeated five times, and the data points (excluding bar graphs) in Fig. 3 and 4 represent the average of those five trials. Error bars correspond to the standard deviations. Below we analyze these errors in details.

\textbf{Errors in Three-Qubit Experiments.} The experimental results shown in Fig. 4 are obtained through the experimental final state of the three-qubit quantum circuit shown in Fig. 2(a).
Since it is not feasible to perform projective measurements in an ensemble NMR system to directly obtain the collapsed states and probabilities, we
reconstructed the quantum state of a three-bit quantum system and applied mathematical post-processing to derive the experimental results. Let $\rho^{exp}_3$ be  the final experimental state of the three-qubit system (comprising the probe P and qubits A and B).
We achieve a fidelity above 0.98 between the measured final states and theoretical predictions.

\textit{Universal Cloner Analysis.} For the universal cloner results shown in Fig. 4(b), the Pauli expectation values are calculated by $$\langle\sigma_\alpha\rangle=\rm{Tr}(\rho_A\sigma_\alpha), \quad \rho_A=\rm{Tr}_{A}{\rho_0}, \quad \rho_{i}=\frac{\rm{Tr}_\text{P}(\rho_3^{exp} P_i)}{p_i}$$ 
where $p_i=\rm{Tr}(\rho_3^{exp} P_i)$ and $P_i=\ketbra{i}$ are defined in the main text. According to the standard error propagation formula, the inaccuracy in $\rho_0$ dominates the uncertainty in the measured Pauli expectation values, leading to mean deviations of approximately $0.018$, $0.014$ and $0.012$ for the three different types of input states, respectively.

\textit{Post-Processing Case.} After post-processing, the results in Fig. 4(b) are calculated as  
$$\langle \sigma_\alpha \rangle=\rm{Tr}(\rho_A\sigma_\alpha), \quad \rho_A=\rm{Tr}_{A}{\rho},\quad \rho=2(p_0\rho_0-p_1\rho_1)=2\rm{Tr}_\text{P}(\rho_3^{exp} P_0)-2\rm{Tr}_\text{P}(\rho_3^{exp} P_1).$$ 
Here, both $\rho_0$ and $\rho_1$ contribute to the overall uncertainty, making the deviation larger than that of the universal cloner case, approximately $0.029$, $0.025$ and $0.015$ for the three different types of input states, respectively.

\textbf{Errors in Four-Qubit Experiments.} The experimental results shown in Fig. 3 are derived from the final four-qubit state $\rho_4^{exp}$, measured at the conclusion of the quantum circuit in Fig. 3(a). Its fidelity to the theoretical prediction is above 0.96, slightly below that of $\rho_3^{exp}$ because the initial state preparation here is more intricate. We first set $\theta=\pi/2$ to obtain the experimental Choi state of VQB, $$C_\mathcal{B}=2(p_0 C_{\mathcal{B}^+}-p_1 C_{\mathcal{B}^-}),\quad C_\mathcal{B}^+=\rho_0(\pi/2),\quad C_\mathcal{B}^-=\rho_1(\pi/2),$$  where $\rho_{0}(\pi/2)$ ($\rho_{1}(\pi/2)$) represents the reduced density matrix of the other three qubits when the control qubit is projected onto $\ketbra{0}$ ($\ketbra{1}$),  as shown in Fig. 3(b). 

The experimental Choi state of the CPTP map $\mathcal{N}_\theta$  and its associated probability $p_0$ for different $\theta$ can be derived from the experimental $\rho_4^{exp}$ as follows: $$C_{\mathcal{N}_\theta}=\rho_0(\theta)=\frac{\rm{Tr}_\text{P}(\rho_4^{exp} P_0)}{p_0}, \quad p_0=\rm{Tr}(P_0\rho_4^{exp}).$$ The uncertainty in $p_0$ originates from the inaccuracy of $\rho_4^{exp}$, yielding a mean deviation of about 0.02. Additionally, the trace distance between the CPTP map $\mathcal{N}_\theta$  and the VQB map $\mathcal{B}$, , defined as $$||C_{\map{B}} -C_{\map{N}_{\theta}}||_1=||\rho_0(\theta)-p_0(\pi/2)\rho_0(\pi/2)-p_1(\pi/2)\rho_1(\pi/2)||_1,$$ is affected by errors in both $\rho_4^{exp}(\theta)$  and $\rho_4^{exp}(\pi/2)$. These results in an average error bar of approximately 0.039.
\end{document}